%Paper: hep-th/9406092
%From: stephany@shaddam.usb.ve (Jorge Stephany Ruiz)
%Date: Tue, 14 Jun 1994 14:53:02 -0400 (GMT-0400)

\input amstex
\input vanilla.sty
\magnification=1200
\TagsOnRight
\nopagenumbers
\headline={\ifnum\pageno=1\hfil\else\hss\tenrm -- \folio\ --\hss\fi}

\def\wh{\widehat}
\def\wt{\widetilde}

\def\noi{\noindent}
\def\ep{\varepsilon}

\TagsOnRight
\mathsurround=1pt
\line{Preprint {\bf SB/F/93-210}}
\hrule
\vskip 1cm
\def\ep{\varepsilon}
\centerline{\bf GAUGE INVARIANCE AND SECOND CLASS CONSTRAINTS}
\centerline{\bf IN 3-D LINERIZED MASSIVE GRAVITY}
\vskip 1cm
\centerline{P\'{\i}o Jos\'e Arias}
\centerline{and}
\centerline{Jorge Stephany}
\vskip 5mm
\centerline{\it Universidad Sim\'on Bol\'{\i}var, Departamento de F\'{\i}sica}
\centerline{\it Apartado Postal 89000, Caracas 1080-A}
\centerline{\it Venezuela}
\centerline{\it e-mail: stephany{\@}usb.ve, parias{\@}usb.ve}
\vskip 3cm

{\narrower\flushpar{{\bf Abstract}

\vskip 3mm

\noi
A recently introduced approach for the dynamical analysis and quantization of
field theoretical models with second class constraints is ilustrated applied
to linearized gravity in 3-D. The canonical structure of two different models
of linerized gravity in 3-D, the intermediate and the self dual models, is
discussed in detail. It is shown that the first order self dual model whose
constraints are all second class may be regarded as a gauge fixed version of
the second order gauge invariant intermediate model. In particular it is
shown  how to construct the gauge invariant hamiltonian of the intermediate
model starting from the one of the self dual model. The relation with the t
opologically massive linearized gravity is also discussed.

\vskip 3cm

\hrule
\bigskip
\centerline{\bf UNIVERSIDAD SIMON BOLIVAR}

\newpage
\noindent
{\bf I. Introduction}
\vskip 3mm

The dynamical analysis and quantization of constrained systems [1] has
been actively studied for many years, but still presents unsolved problems
which represent serious obstacles to extract the physical content of different
models. Some of these problems emerged from the very involved, and sometimes
bizarre constraint structure of some of the systems of physical interest. But
some others are of a very fundamental nature and relate with the basis of our
understanding of, for example, what is a gauge symmetry. This is the case with
the very old problem of the quantization of systems with infinitely many
degrees of freedom and second class constraints [1]. For these systems
at odds with what occurs when there are  only a finite number of degrees of
freedom the Dirac Bracket construction in the operational approach or the
Senjanovic-Fradkin path integral [2] does not represent a practical solution
in most of the cases. The reason is that in general, as occurs for example in
Superparticle and Superstring models [3], the inversion of the Poisson Bracket
matrix of the constraints leads to non-local cumbersome expressions. After the
development of the Batalin, Fradkin and Vilkoviski (BFV) method [4] for
models with  first class constraints, one natural approach to follow for
second class systems is to search for an associate canonically equivalent
model, with only first class constraints. This idea was explored in some
recent papers which addresed this problem [5][6][7][8].

In Ref. [5], a method was presented for the construction of an enlarged phase
space where the original second class constraints have first class
counterparts. The associated BRST charge in the enlarged phase space, in terms
of an auxiliary operator algebra was also discussed. In Ref. [6] some of the
ideas of Ref. [5] were developed further leading to a much simpler formulation
of the associate first  class model in the enlarged phase space. Some
aspects of the reduction to the original degrees of freedom were cleared up
and in particular the equivalence of the original and enlarged models in the
path integral formalism was established. Applied to the
Casalbuoni-Brink-Schwarz superparticle this allow a canonical covariant
construction of the BRST operator [6].

In Ref. [7] and [8] a somewhat more direct approach for this problem was
presented. There, a caraterization of the models with second class constraints,
which can be viewed as gauge fixed versions of gauge invariant systems was
tried. If one can identify, out of the original second class constraints
$\theta_i$, a constraint $\varphi$ with vanishing Poisson bracket with itself
and with all the other constraints but one (say $\chi$) one observe that the
Senjanovic-Fradkin [2] measure split in the form:
$$
0\ne det{\{\theta_i,\theta_i\}}^{1/2}=det \{\varphi ,\chi \}
det{\{\psi_i,\psi_j\}}^{1/2}\tag 1.1
$$
Here $\psi_i$ stand for the remaining constraints. This is the measure adequate
for a gauge theory with gauge transformations generated by $\varphi$, and
gauge fixing condition $\chi$ subjected to the original second class
constraints $\psi_j$. It remains to find the corresponding gauge invariant
hamiltonian. This also can be done but it turns out that at this point
non-local terms may  reappear.

This approach may be of interest also for second class models
for which the Poisson matrix may be handle without much trouble, since it may
shed light on their underlying structure. This is the case with self dual and
topologically massive electrodynamics in 3-D and with the various models of
linearized massive gravity  available in 3-D which we will discuss in this
paper. All these models and many other field theories in 3-D have been studied
intensively in recent years [9][10][11][12][13][14]. The linearized
topologically massive model (TMM) [9] which is a gauge invariant model is
known to have the same spectrum and in this sense to be equivalent to two
other models in 3-D. These are the self-dual model (SDM) and the so called
intermediate model (IM) [10][11]. The IM corresponds also to the
linearization of the curved Vector Chern-Simons gravity action [12]. Both of
them have second class constraints. Moreover they are respectively first
order and second order in space-time derivates at the time that (TMM) is
third order. Finally the IM presents a reduced gauge symmetry respect to the
(TMM) and the (SDM) have no gauge symmetry at all. In Ref. [10] the three
actions were shown to be deducible covariantly from a unique master action.
Although indicative this fact does not established the canonical equivalence
of the systems. There exists examples of systems connected in such fashion
which are not equivalent [15]. In this paper we demostrate the canonical
equivalence of the (SDM) and the (IM). We show that the (IM) is a gauge
theory associated to the (SDM) and we construct explicitly, using the
methods presented in Ref. [8], the gauge invariant hamiltonian of the former
starting from the hamiltonian of the latter.

Before enter these matters, let us  return to our discussion of the
determination of the gauge theory associated to a system with second class
constraints [8]. To consider a slightly more general case suppose that we
have a system subjected to constraints
$$\align
\varphi_i(p,q) & = 0\  \  \  \ i=1,\cdots N \tag 1.2a\\
\chi^i(p,q) & = 0\  \  \  \ i=1,\cdots N \tag 1.2b\\
\psi_j(p,q) & = 0\  \  \  \ j=1,\cdots 2M \tag 1.2c\\
\endalign
$$
satisfying
$$\align
\{ \varphi_i,\varphi_j\} & = C_{ij}{}^k\varphi_k \tag 1.3a \\
\{ \varphi_i,\psi_j\} & = D_{ij}{}^k\varphi_k \tag 1.3b \\
det\{ \varphi_i,\chi_j\} & \ne 0 \tag 1.3c
\endalign
$$

Then (1) generalize in the obvious way with the $\varphi_i$ the first class
constraints, $\chi^i$ the gauge fixing conditions and  $\psi_j$ the
truly second class constraints. Let $H_0$ be the hamiltonian of the system.
The associated gauge invariant hamiltonian must be of the form [8]:
$$
\wt{H}=H_0+\alpha^i\varphi_i+\beta_i\chi^i+\beta_{jk}\chi^j\chi^k+
\beta_{jkl}\chi^j\chi^k\chi^l+\cdots \tag 1.4
$$
over the manifold defined by (1.2c)
The coefficients in this expansions are determined by the condition
$$
\{\wt{H},\varphi_i\} = V_i{}^i\varphi_j\tag 1.5
$$
and are of the form [8]
$$\align
\beta_j &=-\{\chi^j,\varphi_i\}^{-1}\{H_0,\varphi_i\}\\
2\beta_{jk} &=-\{\chi^k,\varphi_i\}^{-1}\{\beta_j,\varphi_i\}\\
3\beta_{jkl} &=-\{\chi^l,\varphi_i\}^{-1}\{\beta_{jk},\varphi_i\}\\
\vdots & \\
V_j{}^i &=-\{\alpha^j,\varphi_i\}+\alpha^jC_{ki}^j\tag 1.6
\endalign
$$

A similar construction allows to obtain a gauge invariant version of any
object in the theory. The appearence of the factor
${\{\chi_j,\varphi_i\}}^{-1}$ in (1.6) implies as mentioned that this
approach may also have problems with non-localities.
\vskip .5cm
\noindent
{\bf II. Self dual and topollogically massive spin 1 in 3-D}
\vskip 3mm

To motivate our presentation to the spin 2 models in section III let us review
briefly in this section the case of the spin 1 as discussed in Ref. [8]. In
$(2+1)D$ massive spin one exitations may be described by the self dual or the
topologically massive models defined respectively by
$$\align
S^{SD} & = \frac{m}{2}<mB_\mu B^\mu-\ep^{\mu\alpha\rho}B_\mu
\partial_\alpha B_\rho >\tag 2.1\\
S^{TM} & = \frac{1}{2}<-F_\mu F^\mu+mF^\mu A_\mu>\  \ ; \  \
F^\mu =\ep^{\mu\alpha\rho}\partial_\alpha A_\rho\tag 2.2
\endalign
$$

For the self-dual model $B_0$ is a Lagrange multiplier. Replacing the
associated constraint one has:
$$\align
H^{SD}_0 & = <\frac{m^2}{2}B_iB_i+\frac{1}{2}(\ep_{ij}\partial_iB_j)^2>
\tag 2.3a\\
P_i & =-\frac{m}{2}\ep_{ik}B_k\tag 2.3b
\endalign
$$

The second class constraints (2.3b) may be replaced by the following
rotationally invariant ones:
$$\align
\varphi & = \partial_iP_i+\frac{m}{2}\ep_{ik}\partial_iB_k\simeq 0
\tag 2.4a\\
\chi & = -\ep_{ij}\partial_iP_j+\frac{m}{2}\partial_kB_k\simeq 0
\tag 2.4b
\endalign
$$

Imposing canonical Poisson brackets for $P_j$ and $B_k$ we have
$$
\{\varphi (x),\chi (y)\}=-m\partial^x_j\partial^y_j\delta^2(x-y)\tag 2.5
$$

We identify $\varphi (x)$ as the generator of the gauge transformations and
$\chi (x)$ as the gauge fixing condition. The  gauge invariant hamiltonian is
$$
\wt{H}=H_0^{SD}+<\beta_1(x,z_1)\chi (z_1)>+
<\beta_2(x_1,z_1,z_2)\chi (z_1)\chi (z_2)>+<\alpha (x)\varphi (x)>\tag 2.6
$$

{}From (1.6) we have
$$\align
\beta_1(x_1,z_1) & = m K(x_1-z_1)\partial_iB_i(x_1)\tag 2.7a\\
\beta_2(x_1,z_1,z_2) & = -\frac{1}{2}K(x_1-z_2)\delta^2(x_1-z_1)\tag 2.7b
\endalign
$$
where
$$
\Delta^x K(x)=-\delta^2(x)\tag 2.8
$$

Using the transverse + longitudinal (T+L) decomposition
$$
B_i(x) = \partial_i B^L+\ep_{ij}\partial_jB^T\tag 2.9
$$
in (2.6) we get the family of gauge invariant hamiltonians
$$
\wt{H} = H_0^{SD}+<\frac{3m^2}{4}B^L\Delta B^L+\frac{m}{2}B^L\Delta
P^T-\frac{1}{2}P^T\Delta P^T>+<\alpha (x)\varphi (x)>\tag 2.10
$$
If one choose
$$
\alpha = -\frac{3m}{4}B^T-\frac{1}{2}P^L\tag 2.11
$$
we get finally
$$
\wt{H}=H^{TM}=\frac{1}{2}P_kP_k+\frac{1}{2}\ep_{ij}P_jB_i+
\frac{m^2}{8}B_kB_k+\frac{1}{2}(\ep_{ij}\partial_iB_j)^2\tag 2.12
$$
This hamiltonian is local and is exactly the one of the
topologically massive model (2.2) (see [16] for example) and hence, this
construction shows the canonical equivalence of the two spin one models.

\vskip 5mm
\noindent
{\bf III. The self dual and the intermediate actions for spin 2 in 3-D}
\vskip 3mm

In this section applying the methods described in the previous ones we
establish the canonical equivalence between the IM and the SDM. The action
for the IM is given by:
$$
S^I=\frac{1}{2}<\ep^{\nu\alpha\beta}\partial_\alpha h_{\beta\rho}
\ep^{\rho\mu\sigma}\partial_\mu h_{\alpha\nu}>-\frac{1}{4}
<(\ep^{\alpha\beta\gamma}\partial_\beta h_{\gamma\alpha})^2>+\frac{m}{2}
<h_{\rho\sigma}\ep^{\rho\mu\nu}\partial_\mu h_{\nu}^{\sigma}>\tag 3.1
$$
where $h_{\alpha\beta}$ is an arbitrary second rank tensor. This action is
invariant under the transformations
$$
\delta h_{\alpha\beta}=\partial_\alpha \xi_\beta \tag 3.2
$$

Let $\pi_{ij}$ be the conjugate momenta to $h_{ij}$. We introduce the
decompositions
$$\align
h_{ij} &= \ep_{ij}\wh{h}+h_{ij}^s\  \ , \  \ h_{ij}^s=h_{ji}^s
\tag 3.3a\\
\pi_{ij} &= \frac{1}{2}\ep_{ij}\wh{\pi}+\pi_{ij}^s\  \ , \  \
\pi_{ij}^s=\pi^s_{ji} \tag 3.3b
\endalign
$$
(we use $\eta_{ij}=(+--),\ep^{012}=1,\ep_{ij}=\ep_{0ij}$). In
general a superscript $s$ will denote a spatial symmetric tensor.

Using these variables one can shown easily that the action does not depends on
$\dot{h}_{00}$, $\dot{\wh{h}}$ and $\dot{h}_{0j}$ and depends lineary on
$\dot{h}_{l0}$. Hence the system is subject to the primary constraints
$$\align
\pi_{00} &\approx 0,\pi_{0j} \approx 0, \tag 3.4\\
\wh{\pi} &\approx 0  \tag 3.5a\\
\Omega_l &\equiv -\ep_{lk}\ep_{ij}\partial_i
h^s_{jk} -\frac{m}{2}\ep_{li}h_{i0} +\pi_{l0}\approx 0\tag 3.5b
\endalign
$$

Imposing the conservation of (3.4) and (3.5) since $h_{00}$, ${\wh{h}}$ and
$h_{0j}$ are lagrange multipliers, generate the secondary constraints
$$\align
\theta_0 &\equiv \ep_{ij}\ep_{kl}\partial_l\partial_ih^s_{jk}-
m\ep_{ij}\partial_i h_{j0}\approx 0\tag 3.6a\\
\theta_k &\equiv -\partial_j\pi^s_{jk}-\frac{3m}{4}\ep_{j l}
\partial_lh_{j k}^s+\frac{m}{4}
\ep_{j k}\partial_l h^s_{j l}+\frac{1}{2}\ep_{lk}
\ep_{ij}\partial_i\partial_lh_{j0}\approx 0\tag 3.6b\\
\psi &\equiv 2m\wh{h}-\pi_{ll}\approx 0\tag 3.7
\endalign
$$

Inspection of the Poisson matrix of the constraints shows that there are
 three first class  constraints given by
$$
\align
\Theta_0 &=\theta-\partial_i\Omega_i=-\partial_i\pi_{i0}-\frac{m}{2}
\ep_{ij}\partial_ih_{j0}\tag 3.8a\\
\Theta_k &=\theta_k + \frac{1}{2}\ep_{kl}\partial_l\wh{\pi} \tag 3.8b
\endalign
$$
which we take instead of (3.6). Equations (3.4), (3.5), (3.6b), (3.7) and
(3.8)  give the complete set of constraints of the systems. The
constraints (3.4)  are first class and appear, to allow writing the action
in a covariant  manner. They are analogous to $\pi_0\approx 0$ in
electrodynamics.  $\wh{\pi},\Omega_l$ and $\psi$ are  second class
constraints.  $\Theta_A$ $(A=0,1,2)$ as stated above are first class and
are the truly  gauge generators of the model. The 18-dimensional phase
space is then reduced  by 4 second class constraints, 6 first class
constraints are 6 gauge fixing  conditions, to the 2 degrees of freedom
needed to describe the single exitation of the system.

After substituting the constraints, the hamiltonian for this model is
given by:
$$\align
H^I=< &-\frac{1}{4}\ep_{ij}\partial_ih_{j0}\ep_{kl}\partial_k h_{l0}
-\frac{m^2}{16}\ep_{ik}\ep_{jl}h^s_{ij}h^s_{kl}
+\frac{m^2}{16}h^s_{ij}h^s_{ij}\\
&-\frac{m}{2}\ep_{ij}h^s_{ik}\pi_{jk}^s -\frac{1}{4}\ep_{ik}
\ep_{jl}\pi_{ij}^s\pi_{kl}^s +
\frac{1}{4}\pi_{ij}^s\pi_{ij}^s>\tag 3.9 \endalign
$$
The gauge transformation laws for $h_{j0}$ and $h^s_{ij}$ are easily obtained
and take the expected form
$$\align
\delta h_{j0} &= \int d^2 x\{ h_{j0}, \xi_A (x)\Theta_A (x)\}=\partial_j
\xi_0 \tag 3.10a\\
\delta h_{ij} &= \int d^2 x \{ h_{ij},\xi_A (x) \Theta_A (x)\} = \frac{1}{2}
(\partial_i\xi_j+ \partial_j\xi_i )  \tag 3.10b\\
\delta \wh{h} &= \int d^2 x \{ \wh{h},\xi_A (x) \Theta_A (x)\}= \frac {1}{2}
\ep_{ij}\partial_i \xi_j \tag 3.10c
\endalign
$$

Let us turn to the SDM. It is defined by the action
$$
S^{SD}=<-\frac{m}{2}\ep^{\alpha\beta\gamma}W_{\alpha\rho}\partial_\beta
W_{\gamma}^{\rho}-\frac{m^2}{2}(W_{\mu\nu}W^{\nu\mu}-W_\rho^\rho
W_\sigma^\sigma)> \tag 3.11
$$
in terms of the second rank tensor $W_{\mu\alpha}$. Let $P_{\mu\alpha}$ be
the conjugate momenta. We define the symmetric and antisymmetric components of
$W_{ij}$ and $P_{ij}$ as in (3.3). Since (3.11) is first order in time
derivates it is not possible to resolve the momenta in terms of the
velocities. There appear nine primary constraints, one associated to each
momenta. We have
$$\align
\varphi_{ij}^s &\equiv P^s_{ij}-\frac{m}{4}(\ep_{ik}W^s_{kj}+
\ep_{kj}W^s_{ki})+{m}\delta_{ij}\wh{W}\approx 0\tag 3.12a\\
\wh{P} &\approx 0\tag 3.12b\\
\varphi_i &\equiv {P}_{i0}+\frac{m}{2}\ep_{ij}W_{j0}\approx 0\tag 3.12c\\
P_{00} &\approx 0\approx P_{0j}\tag 3.13\\
\endalign
$$

The constraints (3.13) are first class and so they can be used to remove six
degrees of freedom. Conservation of (3.13) led to the secondary constraints
$$\align
\zeta &\equiv m W^s_{ll}+\ep_{ij}\partial_iW_{j0}\approx 0\tag 3.14\\
\zeta_k &= mW_{k0}+\ep_{ij}\partial_iW_{jk}^s-\partial_k\wh{W}\approx 0
\tag 3.15
\endalign
$$
This gives twelve constraints and three gauge fixing conditions restricting
the 18 dimensional phase space. Looking for terciary constraints one verifies
that conservation of
$$
\lambda \equiv \wh{P}+\zeta -\frac{\partial_i\varphi_i}{m}\tag 3.16
$$
which we take instead of $\zeta$  implies the last constraint
$$
\wh{W} \approx 0\tag 3.17
$$

The hamiltonian for the SDM is simply
$$
H^{SD}=<\frac{m^2}{2}W_{ij}^sW_{ij}^s-\frac{m^2}{2}W_{ll}^sW_{jj}^s>\tag
3.18 $$

The momenta $P_{0j}$ and $P_{00}$ and the lagrange multipliers $W_{00}$ and
$W_{0j}$ play in this model the same role that $\pi_{oj}$, $\pi_{00}$,
$h_{00}$ and $h_{0j}$ in the IM and we have not to be worried further about
them. Eq. (3.12b) and (3.17) set the canonically conjugate pair
$(\wh{P},\wh{W})$ to zero. We verify explicitly that $\varphi_{ij}^s$,
$\wh{P}$, $\varphi_i$, and $\lambda$ are second class. To show the canonical
equivalence of the SDM and the IM our strategy in what follows consists of:
$i)$ To recognize in the set of constraints of the SDM ((3.12), (3.13),
(3.14), (3.15) and (3.17)) the first and second class constraints of the IM
((3.5), (3.6b), (3.7), (3.8)), $ii)$ To relate the remaining constraints of
the SDM with gauge fixing conditions in the IM and $iii)$ To show that the
gauge invariant hamiltonian of the IM may be written in the form given
schematically by (1.4).

Using (3.17) the first class constraints (3.8) of the IM may be recovered in
the form
$$\align
\Theta_0|_{h=W;\ \pi =P} &=-\partial_i \varphi_i\tag 3.19a\\
\Theta_k|_{h=W;\ \pi =P} &=-\partial_j\varphi^s_{kj}-\frac{1}{2m}\ep_{kj}
\partial_j(\partial_l\varphi_{l}+ m \lambda-2m\wh{P})+m\partial_i\wh{W}
\tag 3.19b
\endalign
$$

The second class constraints (3.5) and (3.7) of the IM, in turn, may be
reconstructed in the form:
$$\align
\wh{P}|_{h=W;\ \pi =P} &=\wh{P} \tag 3.20a\\
\Omega_i|_{h=W;\ \pi =P} &=\varphi_i-\ep_{ij}\partial_i\wh{W}\tag 3.20b\\
\psi|_{h=W;\ \pi =P} &=4m\wt{W}-\varphi_{ll}\tag 3.20c
\endalign
$$

This completes the reconstruction of the constraints of the IM. The
physical sub-manifold of the SDM may be then described by (3.19), (3.20)
and a set of three other independent constraints also constructed by
combinations of $\varphi_{ij}^s$, $\wh{P}$, $\varphi_i$, $\zeta_k$,
$\lambda$ and $\wh{W}$ which are to be interpreted as gauge fixing
conditions asociated to (3.19). We are free to choose any such set with the
only requirement of being independent of (3.19) and (3.20). It can be seen
that this independent  set is expanded by $\lambda$, $\varphi_{ii}$ and
$\ep_{ij}$ $\partial_i$  $\varphi_j$. Good choices  $\chi_A$ for this gauge
fixing
conditions are
$$\align
\chi_0 &=\ep_{ij} \partial_i \varphi_j+\frac{1}{4m}\Delta\varphi_{ii}
\tag 3.21\\
\chi_i &=-\partial_i\lambda+\frac{1}{2}\ep_{ij}\partial_j \varphi_{kk}
\tag 3.22
\endalign
$$
They satisfy
$$
\{\chi_A(x),\theta_B(y)\}=m\delta_{AB}\partial_j^x\partial_j^y\delta (x-y)
\tag 3.23
$$
in analogy with the vectorial case  (2.5).

Let us discuss now the relation between $H^I$ and $H^{SD}$. An explicit
computation shows that on the manifold defined by (3.19), (3.20), (3.21) and
(3.22) we have
$$
H^I|_{h=W;\ \pi =P}=H^{SD}\tag 3.24
$$

Equation (3.24) is a necessary condition for $H^I$ to have the structure given
by (1.4). To complete the proof of the canonical equivalence between the SDM
and the IM let us compute explicitly the coefficients in the expansion (1.4).
which in this case simplify to
$$
\align
H^I= H^{SD} &+<\alpha_A\Theta_A+\alpha_P\wh{P}+
\alpha_\psi\psi+\alpha_i\Omega_i>\\
 &+<\beta_A\chi_A>+<\beta_{AB}\chi_A\chi_B> \tag 3.25
\endalign
$$
Here no additional terms are needed because the hamiltonian is quadratic
in the fields. Then, following the general method described in the
introduction, we obtain
$$
\align
\beta_0(x,z) & =0\tag 3.26\\
\beta_k(x,z) & =-\mu k(x-z)(\ep_{kl}\ep_{ij}\partial_iW^{(s)}_{jl}
\tag 3.27
\endalign
$$
and the only non vanishing $\beta_{AB}$ is
$$
\beta_{kl}(x,z_1,z_2)=-\frac{1}{4}K(x-z_1)\ep_{ki}\partial^{z_2}_i
\ep_{lj}\partial^{z_2}_jK(z_2-z_1)\tag 3.28
$$
{}From (3.9),(3.18) and (3.25) one can compute the expression for the
$\alpha$'s. This can be done for example factorizing the constraints in
(3.9) and comparing with (3.18). Then one can see that the $\beta$'s
appear naturally and we obtain:
$$\align
<\alpha_A\Theta_A &+\alpha_P\wh{P}+
\alpha_\psi\psi+\alpha_i\Omega_i>=\\
 &<\Theta_i(-\Delta )^{-1}[\partial_iP_{kk}-\partial_kP_{ki}-\frac{1}{2}
\ep_{ik}\partial_k\wh{P}-\frac{m}{4}\ep{kl}\partial_kW_{li}+\\
&\ \ \ \ \ \ \ \ \ \ \ \ \ \ \ \ \ \
-\frac{3m}{4}\ep_{ik}\partial_kW_{ll}-\frac{1}{2}(\Delta \delta_{ik}-
\partial_i\partial_k)W_{k0}]+\\ &+\Theta_0[\frac{1}{2m}\wh{P}+(-\Delta
)^{-1}(\delta_{kl}\Delta -\partial_k \partial_l)W_{kl}]+\\
&+\psi[-\frac{3}{16}P_{ii}-\frac{m}{8}\wh{W}-\frac{m}{2}(-\Delta
)^{-1}\ep_{ik} \partial_i\partial_lW_{lk}]+\\
&+\wh{P}[\frac{1}{2m}\partial_iP_{i0}+\frac{1}{4}\wh{P}-\frac{1}{4}\ep_{ik}
\partial_iW_{k0}+2m(-\Delta )^{-1}(\Delta
\delta_{ik}-\partial_i\partial_k) W_{ik}]>\tag 3.29
\endalign
$$
Introducing (3.29) in (3.25) we have an identity which establishes the
canonical equivalence of SDM and IM and our interpretation of the SDM as
an explicitely covariant gauge fixed version of the IM.

\vskip 5mm
\noindent
{\bf IV The full gauge invariant extension for $H^I$ }
\vskip 3mm

In the last section, there were 4 second class constraints, represented by
$\psi$, $\wh{\pi}$ and $\Omega_l$ ((3.5), (3.7)), that did not play any role
in the proof of the canonical equivalence between the $SDM$ and the $IM$.
They could be used, futher, to obtain a full gauge invariant form of $H^I$,
and expedite the construction of the effective action for the $IM$, and
consequently for $SDM$.

Let us rename and redefine this set of constraints as
$$\align
\Theta_3 &\equiv \Delta\wh{\pi} \tag 4.1,a \\
\Theta_4 &\equiv \partial_i\Omega_i=\partial_i\pi_{i0}-\frac{m}{2}\ep_{ij}
\partial_ih_{j0}-(\delta_{ij}\Delta -\partial_i\partial_j)h_{ij}\tag 4.1,b\\
\chi_3 &\equiv \frac{1}{2}\psi -\frac{1}{2m}\ep_{ij}\partial_i\Omega_j=
m^2\wh{h}-\frac{1}{2}\pi_{ii}^{(s)}-\frac{1}{2m}\ep_{ij}\partial_i\pi_j-
\frac{1}{2m}\ep_{ij}\partial_i\partial_kh_{kj}^{(s)}-\frac{1}{4}\partial_i
h_{i0}\tag 4.1,c\\
\chi_4 &\equiv -\ep_{ij}\partial_i\omega_j=-\ep_{ij}\partial_i
\pi_{j0}-\frac{m}{2}\partial_ih_{i0}-\ep_{ij}\partial_i\partial_k
h_{jk} \tag 4.1,d
\endalign
$$
where we are sugesting to take $\Theta_3$ and $\Theta_4$ as first
class constraints and the $\chi$'s as gauge fixing conditions. Their Poisson
matrix is
$$
\{\Theta_{A'}(x),\Theta_{B'}(y)\} =m\delta_{A'B'}\partial_j^x\partial_j^y
\delta(x-y)\tag 4.2
$$

Let $\wt{H}^I$ be the full gauge invariant form of $H^I$. Like in (1.4) we
impose that
$$
\wt{H}^I=H^I+<\alpha_{A'}\Theta_{A'}>+<\beta_{A'}\chi_{A'}>+
<\beta_{A'B'}\chi_{A'}\chi_{B'}>\tag 4.3
$$
where $A'=3,4$. Following the same procedure presented, we obtain
$$\align
\beta_3(x,z) &=0\tag 4.4,a\\
\beta_4(x,z) &=\frac{1}{m}K(x-z)[(\delta_{ij}\Delta -\partial_i\partial_j)
\pi_{ij}(x)+\frac{1}{2}\Delta_x\pi_{ii}(x)+\frac{m}{2}\ep_{ij}
\partial_i\partial_kh_{kj}(x)]\tag 4.4,b\\
\beta_{33}(x_1,z_1,z_2) &=\beta_{34}(x_1z_1,z_2)=\beta_{43}(x_1z_1,z_2)=0
\tag 4.4,c\\
\beta_{44}(x_1,z_1,z_2)&=\frac{-1}{4m^2}\delta(x-z_1)\delta(x-z_2)\tag 4.4,d
\endalign
$$

As we pointed out before the $\alpha$'s in (1.4), (3.24)
and (4.3) remain  unfixed. For a particular choice they
may give a canonical connection with the TMM. Nevertheless since this
model is written in terms of third derivatives its canonical structure
should be analyzed with special care [17] in order to establish such
equivalence. This would be done elsewhere [18].

\newpage \item{}{\bf References} \vskip 3mm

\item{[1]}P. A. M. Dirac, {\it Lectures on Quantum Mechanichs} Belfer
Graduate School of Science, Yeshiva University, New York (1964).
\item{[2]}G. Senjanovic, Ann. Phys. {\bf 100} (1976) 227.
\item{}E. S. Fradkin, {\it Acta Universitatis Wratisaviensis} {\bf N207}
Proc. Xth Wintre School of Theoretical Physics in Karpacz (1973) p93.
\item{[3]}M. B. Green and J. Schwarz, Phys. Lett. {\bf B109} (1983) 399.
\item{}L. Brink and J. Schwarz, Phys Lett {\bf B100} (1981) 287.
\item{}R.Casalbuoni, Nuov Cim {\bf A33} (1976) 432.
\item{[4]}E. S. Fradkin and G. A. Vilkovisky, Phys. Lett. {\bf B55} (1975)
224; CERN report TH-2332 (1977).
\item{}I. Batalin and E. Fradkin, Phys. Lett. {\bf B122} (1983) 157; Phys.
Lett. {\bf B128} (1983) 307; An Inst Henri Poincar\'e {\bf 49} (1988) 215.
\item{[5]}I. A. Batalin and E. S. Fradkin, Phys. Lett. {\bf B180} (1983)
157; Nucl. Phys. {\bf B279} (1987) 514.
\item{}I. A. Batalin, E. S. Fradkin and T. E. Fradkina, Nucl. Phys. {\bf
B314} (1989) 158.
\item{[6]}A. Restuccia and J. Stephany, Phys. Lett.{\bf B305} (1993) 348;
Phys. Rev.{\bf D47} (1993) 3437.
\item{[7]}K. Harada and H.Mukaida, Z. Phys {\bf C48} (1990) 151.
\item{[8]}R. Gianvittorio, A. Restuccia and J. Stephany, Mod. Phys. Lett.
{\bf A6} (1991).
\item{[9]}R. Jackiw, S. Deser and S. Templeton, Ann. Phys. {\bf 140}
(1982)  372.
\item{[10]}C. Aragone and A. Khoudeir, Phys. Lett. {\bf B173} (1986) 141;
Quantum Mechanics of fundamental systems p 86. Ed. C. Teitelboin Plenum Press,
New York (1987).
\item{[11]}S. Deser and J. Mc Carthy, Phys. Lett. {\bf B246} (1990) 441;
Phys. Lett. {\bf B248} (1990) 248.
\item{[12]}C. Aragone, P. J. Arias and A. Khoudeir, to appear in Nuovo
Cimento {\bf B}.
\item{}C. Aragone, P. J. Arias and A. Khoudeir, to appear in in the
Proceedings of SILARG VIII, World Scientific Publishing Brazil (1993).
\item{[13]}P. K. Townsend, K. Pilch and P. van Nieuwenhuizen, Phys. Lett.
{\bf B136} (1984) 38; {\bf B137} (1984) 443.
\item{}S. Deser and R. Jackiw, Phys. Lett. {\bf B139} (1984) 371.
\item{[14]}S. Deser and J. Kay, Phys. Lett. {\bf B120} (1983) 97
\item{}C. Aragone and J. Stephany, Class and Quant. Grav. {\bf 1} (1984)
265. \item{}C. Aragone and S.Deser, Class and Quant. Grav.{\bf 1} (1984)
331. \item{}C. Aragone Class Quant, Grav {\bf 2} (1985) L25.
\item{[15]}C. Aragone and J. Stephany, Phys. Rev. {\bf D34} (1986) 1210.
\item{[16]}J.M.Martinez Fernandez and C.Wotzasek, Z. Phys {\bf C43}(1989)
305.
\item{[17]}V.Aldaya and J.A.de Azcrraga, J.Phys A:Math Gen . {\bf 13}
(1980) 2545; J.Barcelos-Neto and C.P.Natividade Z.Phys C {51} (1991) 313;
 and A.Restuccia Ann Phys (NY) {224} (1993) 1.
\item{[18]}P.Arias and J. Stephany, in preparation.
\bye